\documentstyle[12pt,a4]{article}

\newcommand{\be}{\begin{equation}}
\newcommand{\ee}{\end{equation}}

\title{Classical solutions in the Einstein-Born-Infeld-Abelian-Higgs model}
\author{Y. Brihaye, B. Mercier\\
Facult\'e des Sciences, Universit\'e de Mons-Hainaut,\\
B-7000 MONS, Belgium.}
\date{\ }
\begin{document}
\begin{titlepage}
\maketitle
\thispagestyle{empty}
\begin{abstract}
We consider the classical equations of the
Born-Infeld-Abelian-Higgs model (with and without coupling to gravity)
in an axially symmetric ansatz. 
A numerical analysis of the equations reveals that the (gravitating)
Nielsen-Olesen vortices are smoothly deformed by the Born-Infeld
interaction, characterized by a coupling constant $\beta^2$,
and that these solutions cease to exist at a critical value of $\beta^2$.
When the critical value is approached, the length of the magnetic field
on the symmetry axis becomes infinite.
\end{abstract}
\vfill
\end{titlepage}
\newpage
\section{Introduction}
Recently, various gauge field theories were studied in which the 
dynamics of the gauge fields is governed by a Born-Infeld (BI) lagrangian
\cite{bi}. The relationship between string theory and BI lagrangian
obtained in \cite{tsey1} started an important activity in the topic
which is reviewed e.g. in \cite{pol, tsey2}.

Among the various aspects of BI field theory, the construction 
of classical solutions occupies a large place. Many investigations
were carried out in Abelian \cite{mns} as well as in non-Abelian
\cite{kg,gpss} models. Often, the pattern of classical solutions
is enriched by the occurence of the BI-term. The case of the SU(2)
gauge theory is typical of this type: no static finite
energy solutions exist with the minimal Yang-Mills action
but solutions exist when the BI term is included \cite{kg}.
Lately, various Born-Infeld solitons have been considered
in the context of gravity (see. e.g. \cite{tri,wsk}).

Perhaps the simplest model containing solitons for which the
effects of the BI term may be studied is the Abelian-Higgs model
\cite{no} in which vortex-solutions are available.
Several aspects of BI-vortex solutions have been  considered so far, 
namely the construction of self dual equations from the BI
lagrangian \cite{shir} and the construction of electrically
charged vortex solutions \cite{mns}.
Very interesting applications of the vortex solutions are cosmic
strings \cite{vil} which are believed to 
have played a role in the early phase of the Universe; it is therefore natural to also consider the 
effect of gravitation on the BI-vortex solutions.

Here we consider the Abelian-Higgs-Born-Infeld model coupled to gravity
and we preform a detailed numerical analysis of the equations in the
axially symmetric ansatz.
We find that, when the various coupling constants 
of the Einstein-Abelian-Higgs equations
are such that a magnetically charged
soliton exists (i.e. for $\beta^2 = \infty$ where $\beta^2$ denotes the
BI coupling constant), then the solution is smoothly deformed 
by the BI interaction. When the parameter  $\beta^2$ approaches
a critical value, the length of the
magnetic field of the solution on its
symmetry axis becomes infinite and the solution disappears.

The paper is organized as follows~: in Sect. 2 we describe the
Lagrangian, the ansatz and the classical equations. Several known
solutions and some of their features are summarized in Sect. 3.
Sect. 4 contains our results on the evolution of these 
solutions in presence of the Born-Infeld interaction.

\section{The equations.}
We consider the gravitating 
Born-Infeld-Abelian Higgs model in four dimensions. It is
described by the action \cite{bi,no}
\be
S = \int d^4x \sqrt{-g}{\cal L}
\ee
\be
\label{lagran}
{\cal L} = 
   {1\over 2} D_{\mu} \phi D^{\mu} \phi^{\ast} 
+  \hat \beta^2 (1 - {\cal R}) 
- {\lambda \over 4} (\phi^{\ast} \phi-v^2)^2 
+ {1\over{16 \pi G}} R
\ee
with 
\be
  {\cal R} = \sqrt{1 + \frac{F_{\mu\nu}F^{\mu\nu}}{2 \hat \beta^2}
         -\frac{(F_{\mu\nu}\tilde F^{\mu\nu})^2}{16 \hat \beta^4}
                   }
\ee
where $D_{\mu} = \nabla_{\mu} - i e A_{\mu}$ is the gauge
covariant derivative, $A_{\mu}$ is the gauge potential
of the U(1) gauge symmetry,
$F_{\mu\nu}$ is the
corresponding field strength and $\phi$ is a complex scalar field with vacuum
expectation value $v$. 
The Born-Infeld coupling constant
 is noted $\hat \beta^2$. The geometry is introduced as usual
by means of the Ricci scalar $R$. We use the same
notations as in \cite{verbin}.

\par In the following we study the classical equations
of the above field theory in the static case and within the
 cylindrically symmetric ansatz. The
metric and matter fields are parametrized in terms of four functions of the cylindrical radial variable  $r$  :
\be
ds^2 = N^2(r) dt^2 - dr^2 -  L^2(r) d\varphi^2-N^2(r) dz^2
\ee
\be
\phi = vf(r) e^{i n\varphi}
\ee
\be
A_{\mu} dx^{\mu} = {1\over e} (n-P(r)) d\varphi
\ee
Here $N, L,P,f$ are the  radial functions of $r$, $n$ is an integer
indexing the vorticity of the Higgs field around the $z$-axis.

In the process of establishing the classical equations
in terms of the radial functions  it appears convenient
to define  dimensionless coupling constants ~:
\be
\alpha = \frac{e^2}{\lambda} \ \ , \ \  \gamma = 8\pi G v^2 \ \ , 
\ \ \beta^2 = \frac{\hat \beta^2}{\lambda v^4} \ \ ,
\ee
and to rescale the radial variable $r$ and the function $L(r)$ according to
\be
x = \sqrt{\lambda v^2}r \ \ , \ \ L(x) = \sqrt{\lambda v^2}  L(r)
\ee

\par 
With these redefinitions (and in the units $\hbar = c = 1$)
the diagonal components of the energy-momentum tensor read
\begin{eqnarray}
{\cal T}^0_0 &=& \lambda v^4(\epsilon_s+\epsilon_v+\epsilon_w+u)\\
{\cal T}^r_r &=& \lambda v^4(-\epsilon_s+\epsilon_v-\tilde\epsilon_v+\epsilon_w+u)\\
{\cal T}^{\varphi}_{\varphi} &=& \lambda v^4(\epsilon_s+\epsilon_v-\tilde \epsilon_v-\epsilon_w
+u)\\
{\cal T}^z_z &=& {\cal T}^0_0
\end{eqnarray}
where the different densities are
expressed in the dimensionless variable and
functions~:
\be
\epsilon_s = {1\over 2}(f')^2\quad , \quad \epsilon_w = {P^2f^2\over {2L^2}}\quad , \quad u = {1\over 4} (1-f^2)^2
\ee
\be
\epsilon_v = \beta^2({\cal R}-1)\quad , \quad \tilde \epsilon_v = {(P')^2\over
{\alpha L^2 {\cal R}}}\quad , \quad {\cal R} \equiv \sqrt{1+{(P')^2\over{\alpha\beta^2L^2}}}
\ee
the prime indicates the derivative with respect to $x$.
The corresponding components of the Ricci tensor are given explicitely
e.g. in \cite{verbin}.

Then, after an algebra, the classical equations 
corresponding to the above lagrangian
and ansatz can be shown to be consistent and lead
 to the following
system of four equations for the functions $N,L,P,f$~:
\be
\label{eq1}
{(LNN')'\over{N^2L}} = \gamma \left(
\beta^2 (1-{\cal R}) 
+ \frac{P'^2}{\alpha L^2} \frac{1}{{\cal R}}
-{1\over 4}(1-f^2)^2
\right)
\ee
\be
\label{eq2}
{(N^2L')'\over{N^2L}} = \gamma \left( 
\beta^2 (1-{\cal R}) 
- {P^2f^2 \over{L^2}} 
- {1\over 4} (1-f^2)^2\right)
\ee
\be
\label{eq3}
{L\over{N^2}} (\frac{N^2 P'}{L{\cal R}})' = \alpha f^2 P
\ee
\be
\label{eq4}
{(N^2Lf')'\over{N^2L}} = f(f^2-1) + f {P^2\over {L^2}}
\ee

The problem posed by these differential equations is 
further specified by the following set of boundary conditions
\begin{eqnarray}
\label{cn} N(0) &=& 1\qquad , \qquad N'(0) = 0\\
\label{cl} L(0) &=& 0 \qquad , \qquad L'(0) = 1\\
\label{cp} P(0) &=& n \qquad , \qquad \lim_{x\rightarrow \infty} P(x) = 0 \label{condp} \\
\label{cf} f(0) &=& 0 \qquad , \qquad \lim_{x\rightarrow \infty}f(x) = 1
\end{eqnarray}
which are necessary to guarantee the regularity of the configuration 
on the symmetry axis
and the finiteness of the inertial mass  defined next.

The classical solutions can be characterized, namely,
 by their magnetic field
\be
\label{magnetic}
B(r) = -{1\over{eL(r)}} {dP\over{dr}}(r) = -{\gamma\over{\alpha}}
({e\over{8\pi G}}) {P'(x)\over{L(x)}},
\ee
and by their inertial mass and Tolman mass by unit length \cite{verbin}
\begin{eqnarray}
{\cal M}_{in} &=& \int \sqrt{-g_3} \ {\cal T}^0_0 \ dx_1dx_2\\
{\cal M}_{t_0} &=& \int \sqrt{-g_4} \ 
({\cal T}^0_0 - {\cal T}^r_r- {\cal T}^{\varphi}_{\varphi}- {\cal T}^z_z) \ dx_1dx_2
\end{eqnarray}
which in turn can be reduced to one dimensional integrals~:
\begin{eqnarray}
G{\cal M}_{in} &=& {\gamma\over 8} M_{in}\\
&=& {\gamma\over 8} \int^{\infty}_0 dx NL(\epsilon_s + \epsilon_w + \epsilon_v + u).\\
G{\cal M}_{t_0}&=& {\gamma\over 8} M_{t_0}\\
&=& {\gamma\over 8} \int^{\infty}_0 dx N^2L (\tilde \epsilon_v-\epsilon_v-u).
\end{eqnarray}
In the flat case ($\gamma = 0$) the classical energy $E$
of the solutions
is obtained directly in terms of the inertial mass~: 
$E= (2 \pi v^2) M_{in}$.

\vspace{0.5cm}

\section{Some special cases.}
In this section we briefly 
review a few particular solutions of the system above
and comment on their main properties.

\subsection{Nielsen-Olesen vortices.}

We first consider the equations 
for the case $\gamma=0, \beta=\infty$.
This case corresponds to the flat Maxwell-Higgs system; Eqs. 
(\ref{eq1}), (\ref{eq2}) are 
solved by $N(x) = 1, L(x) = x$. The solutions 
of the remaining equations are embeddings  of the celebrated
two-dimensional Nielsen-Olesen vortices \cite{no} into 
the three-dimensional space. 
They exist for all values of $\alpha$ and $n$
 and possess the following behaviour around the origin
\be
\label{noor}
P(x) = n - \frac{a_0}{2} x^2+O(x^{2+2n})
 \ \ \ , \ \ \ 
f(x) = b_0 x^n + O(x^{n+2})
\ee
and asymptotically
\be
\label{nobord}
P(x) = a_{\infty} e^{-x \sqrt{\alpha}}
\ \ \ , \ \ \ 
f(x) = 1-f_{\infty} e^{-x \sqrt{2}}  \ \ .
\ee
The parameters $a_0, b_0, a_{\infty}, f_{\infty}$ are determined numerically. For fixed $n$,
the classical energy, say $E(\alpha,n)$, of the solution
decreases monotonically when $\alpha$ increases. The
value $\alpha=2$ is very particular because first order equations exist
\cite{taubes} which imply (\ref{eq3}), (\ref{eq4})
(they are usually called self dual equations).
\par In relation to the self dual case, $\alpha=2$, the Nielsen-Olesen
solutions enjoy the following remarkable property :
\begin{eqnarray}
\label{sd}
{1\over n} E(\alpha,n)&>& E(\alpha,1)\qquad {\rm for} \qquad \alpha< 2\nonumber\\
{1\over n} E(\alpha,n) &=& E(\alpha,1)\qquad {\rm for} \qquad \alpha=2\nonumber\\
{1\over n} E(\alpha,n) &<& E(\alpha,1)\qquad {\rm for} \qquad \alpha > 2
\end{eqnarray}
as illustrated by Fig. 1 for $n=1,2$. This supports the idea 
\cite{rebbi} that, in the domain $\alpha > 2$, the n-vortice solutions
(with $n>2$) are stable against decay into $n$
separated solutions with unit vorticity (remembering the exponential decay
(\ref{nobord}) of the solutions under consideration).

\vspace{0.5cm}

\subsection{ The Melvin solution.}
\par Cancelling the Higgs field and the Higgs potential in the initial
lagrangian, we are left with the Einstein-Maxwell lagrangian. The
corresponding equations can be obtained from (\ref{eq1})-(\ref{eq4}) after
an appropriate rescaling of $x$ and $L$~:
\be
x\rightarrow {x\over{\sqrt{\gamma}}}\quad , \quad L \rightarrow
\sqrt{\gamma\over{\alpha}} L
\ee
and setting $\gamma=0$ afterwards
(then Eq.(\ref{eq4}) decouples). 
A solution of these equation was discovered by
Melvin \cite{melvin}. It is namely characterized by 
the following (asymptotic) power law behaviours of the various fields   
\be
N(x\rightarrow \infty) = Ax^{2\over 3}\quad , \quad L(x\rightarrow \infty)
= B x^{-{1\over 3}}\quad , \quad P(x\rightarrow \infty) = Cx^{-{2\over 3}}
\ee
where $A,B,C$ are constants (the numerical evaluation gives $A\approx
2.25, B\approx 0.44)$. The corresponding masses are $M_{in} = M_{t_0}
= 2$.

\vspace{0.5cm}

\subsection{Gravitating vortices} 
\par The solutions corresponding to the Maxwell case
(i.e. $\beta^2 = \infty$)  were studied at length in \cite{verbin,bl}. We just mention
that global solutions (i.e. defined for $x\in [0,\infty])$ exist only on a 
finite domain in the (quarter of) plane $\alpha,\gamma$.
More precisely they exist for
\be
\label{domain}
0 \leq \gamma \leq \gamma_{c}(\alpha,n)
\ee
The function $\gamma_c(\alpha,n)$ increases monotonically with $\alpha$
and, around the self-dual value $\alpha= 2$ it roughly behaves accordingly to
$\gamma_c(\alpha,n) = 2/n + \kappa(\alpha-2) \ , \ \kappa \approx 0.15$.  
In fact, for fixed $\alpha$, there exist two branches of solutions
for the set of $\gamma$ in the domain (\ref{domain}).
Following \cite{verbin} we refer  to the two branches as to  
the string and the Melvin branches. 

For $\gamma > \gamma_{c}(\alpha,n)$, solutions  still do
exist but they are closed 
\cite{verbin}; i.e. they are only defined for $x\in [0,x_{max}]$.
 We will not
be concerned with these type of solutions in the present paper.

\section{Numerical results.}
We solved numerically the system (\ref{eq1})-(\ref{eq4}) for numerous values
of the coupling constants $\alpha,\beta,\gamma$, studying in particular the
response of the three types of solutions mentioned above to
the Born-Infeld interaction. We now discuss then results in the same
order as before.

\subsection{The Nielsen-Olesen-Born-Infeld vortices}
We start by considering $\gamma = 0$. The formula  
(\ref{nobord}) still holds for $\beta < \infty$ and,
as expected, the Nielsen-Olesen solutions are smoothly deformed by the 
Born-Infeld term.  Fig. 2
shows the classical energy 
by means of $M_{in}(\alpha,\beta,n)/n$ as a function of $\beta^2$
for several values of $\alpha$ and $n$. It clearly indicates that the inertial
mass depends only little on $\beta$.
The corresponding values $E_m$ in the Maxwell limit ($\beta^2 = \infty$)
are indicated nearby. The main feature demonstrated by this figure
is that the solutions exist only for a finite range of the parameter $\beta$,
i.e. for  
\be
\beta_{c}(\alpha,n)< \beta < \infty
\ee
confirming the result of \cite{mns}. 
An in-depth analysis about the reasons
for this critical phenomenon is provided by Fig. 3 where
the  numerical evaluation of the parameters $a_0, b_0$ for the solutions are
plotted as functions of $\beta^2$. Our numerical analysis strongly indicates
that 
\be
\label{lim}
\lim_{\beta\rightarrow \beta_{c}} a_0 = \infty
\ee
so that, very likely, no solutions exist for $\beta < \beta_{c}$.
 This critical behaviour is, however, not reflected by the value of the
classical energy and by the parameter $b_0$ which remain finite for 
$\beta\rightarrow \beta_{c}$.
\par Physically, this phenomenon is 
reflected by the magnetic field of the solution. 
Indeed, from the definition (\ref{magnetic}),
 it is immediate to see that $B(0)$
(i.e. the value of the magnetic field on the symmetry axis of the solution) is directly 
related to $a_0$; Eq. (\ref{lim}) can therefore be rephased in saying that the magnetic
field on the z-axis becomes infinite when the critical value of $\beta$ is approached.
 This phenomenon is further illustrated  by Fig. 4 where the
 magnetic function $B(x)$ is compared for several values of $\beta^2$
 in the case $\alpha=1, n=1$. 

Refering to the relations (\ref{sd}), a natural question is to 
analyse how the domain of stability of the
$n>2$-vortices (i.e. $2<\alpha<\infty$) develops with the Born-Infeld interaction. 
An important element of  answer is given by Fig. 1. 
Clearly the effect of the Born-Infeld term is that the 
stable phase appears for lower values of $\alpha$. 
In terms of the Higgs particle mass $M_H = \lambda v$
(remember $\alpha = {e^2\over{\lambda}}$)
this means that the stable phase persists
up to a  higher value of the Higgs particle mass.

\par This result motivates an  analysis
of the effect of the Born-Infeld term in the
Georgi-Glashow model for multi-monopole solutions. 
It is known  that in this model, self dual equations exist
for $\lambda=0$ (i.e. for a massless Higgs particle) and that
multimonopoles solutions are unstable for $\lambda > 0$ \cite{manton}.

It is likely that the Born-Infeld
interaction generates a small domain of the Higgs mass
parameter where the mass of the two-monopole
is lower than twice the mass of the single monopole, thus leading to a stable phase.
Such a phenomenon was observed recently in the gravitating Georgi-Glashow model \cite{hkk}.
Due to gravity, a region of the parameter space exists where the multimonopoles seem to be stable against decay into monopoles.

\subsection{The Melvin-Born-Infeld Solution.}
\par The Melvin solution gets smoothly deformed by the Born-Infeld term. Decreasing
$\beta^2$ we observed that the solution disappears for $\beta^2_{c}\approx 6.4$; again
the phenomenon is related to the fact that the value $B(0)$ tends to infinity for
$\beta^2\rightarrow \beta^2_{c}$. The profiles of $N(x), L(x)$ and of the magnetic
field for the Melvin solution and its deformation for $\beta^2 = 7$ are compared on Fig. 5 (the dotted line for $B$ has $B(0)\approx 40$).

\subsection{The Born-Infeld gravitating strings.}
The study of the equations (\ref{eq1}-\ref{eq4}) for generic values
of $\alpha, \beta, \gamma$ is a vast task. Here, we have attempted to
confirm the features which occured for the flat case. 
We limit ourselves to the case $\alpha = 2$ but we expect that the
results   are qualitatively
the same for the other values of this coupling constant. 
Our numerical analysis of the equations
 confirms that, for $\gamma$ fixed, the solutions
on the string branch as well as the ones on the Melvin branch
terminate at finite values of $\beta^2$.
Fig. 6 shows the domains of existence of the two types of solutions
in the $\beta^2, \gamma$ plane.
The figure clearly reveals that the value $\beta_c^2$, where the 
string branch terminates (the lower curve) varies only little with
$\gamma$ and that is is smaller than its counterpart,
say $\tilde \beta^2_c$, for the Melvin branch
(upper curve).

Taking e.g.  $\alpha=2, \gamma=1$ we find
\be
     \beta^2_{c} \approx 0.4   \ \ \ , \ \ \ 
      \tilde \beta^2_{c} \approx 5.1
\ee
respectively for the solution of the string and of the Melvin branches.
The inertial mass deviates only slightly from the  exact value
$M_{in}=2$ in the Maxwell limit, the Tolman mass decreases slowly
from $M_{to}\approx 0.62$ in the Maxwell limit to $M_{to}\approx 0.4$ 
for $\beta^2 = \tilde \beta^2_{c}$. 

For both branches, the value $B(0)$ of the magnetic field
on the z-axis becomes very large (probably infinite
but our numerical analysis
does not allow us to be absolutely sure) 
when the critical value of $\beta^2$ is approached.
Other quantities, e.g. the inertial and Tolman masses remain finite.

We finally mention that the string branch 
(which exists for $0\leq \gamma \leq
2$ in the case $\alpha = 2$) extends slightly into the region $\gamma > 2$
when the Born-Infeld term is present.

\section{Final remarks}
To finish, we would like to present a generalization
of the result in \cite{shir} to the gravitating case.
The self-dual potential and the corresponding equations
(i.e.  setting $\lambda=e^2/2$ in Eq.(\ref{lagran}) and $\alpha = 2$ in 
Eqs.(\ref{eq1})-(\ref{eq2})) can be replaced by a  $\beta^2$-depending one
for which first order equations exist which imply the classical (second order) 
axially-symmetric equations. 
In  the reduced  variables, this can be achieved by the following substitution
\be
\label{potsd}
V(f) = \frac{1}{4}(f^2-1)^2 \longrightarrow 
  V_{sd}(f) =  \beta^2 \Biggl(   1- 
\sqrt{ 1 - {1 \over 2  \beta^2} (f^2-1)^2 }
                         \Biggl)
\ee
With this potential the self dual equations read
\be
N(x)=1 \ \ , \ \ L'= 1 + \frac{\gamma}{2}( P(1-f^2) + \theta) \ \  , \ \  
\frac{P'}{L} = \frac{1}{f} \frac{\partial V_{sd}}{\partial f} \ \ , \ \ 
f' = \frac{P \ f}{L}
\ee
and imply the full axially symmetric equations of the theory.
Further exact solutions and algebraic properties 
 can be obtained with this modified theory.

\vskip 2 cm 
{\bf Acknowledgements}\hfill\break
 One of us (Y.~B.) gratefully acknowledges conversations 	
 with B.~Hartmann and with F.A.~Schaposnik.

\newpage
\centerline{Figures Captions}
\begin{description}
\item [\ ] {\bf{Figure 1}}\\
The ratio  $E/n$ is plotted as a function of $\alpha$
for the Maxwell and Born-Infeld vortices and for $n=1,2$. 
\item [\ ] {\bf{Figure 2}}\\
The ratio $E/n$ for the Born-Infeld vortices is plotted
as a function of $\beta^2$ for different values of $\alpha$ and $n$.
On each branch, $E_m$ gives the energy of the Maxwell solution.
\item [\ ] {\bf{Figure 3}}\\
The parameters $a_0,b_0$ entering in the solutions via Eq. (\ref{noor})
are reported as functions of $\beta^2$ for different values of
$\alpha$ and of $n$.
\item [\ ] {\bf{Figure 4}}\\
The profiles of the magnetic field $B(x)$ is presented for
several values of $\beta^2$ approaching the critical value 
$\beta_c^2\approx 0.318$.
\item [\ ] {\bf{Figure 5}} \\
The profiles of the functions $L,N,B$ for Melvin solution
(solid lines) and Melvin-Born-Infeld solution (dotted lines).
On the dotted line $B(0) \approx 42$.
\item [\ ] {\bf{Figure 6}} \\
The domain of the gravitating string and Melvin branches of solutions
for $\alpha = 2$.

\end{description}

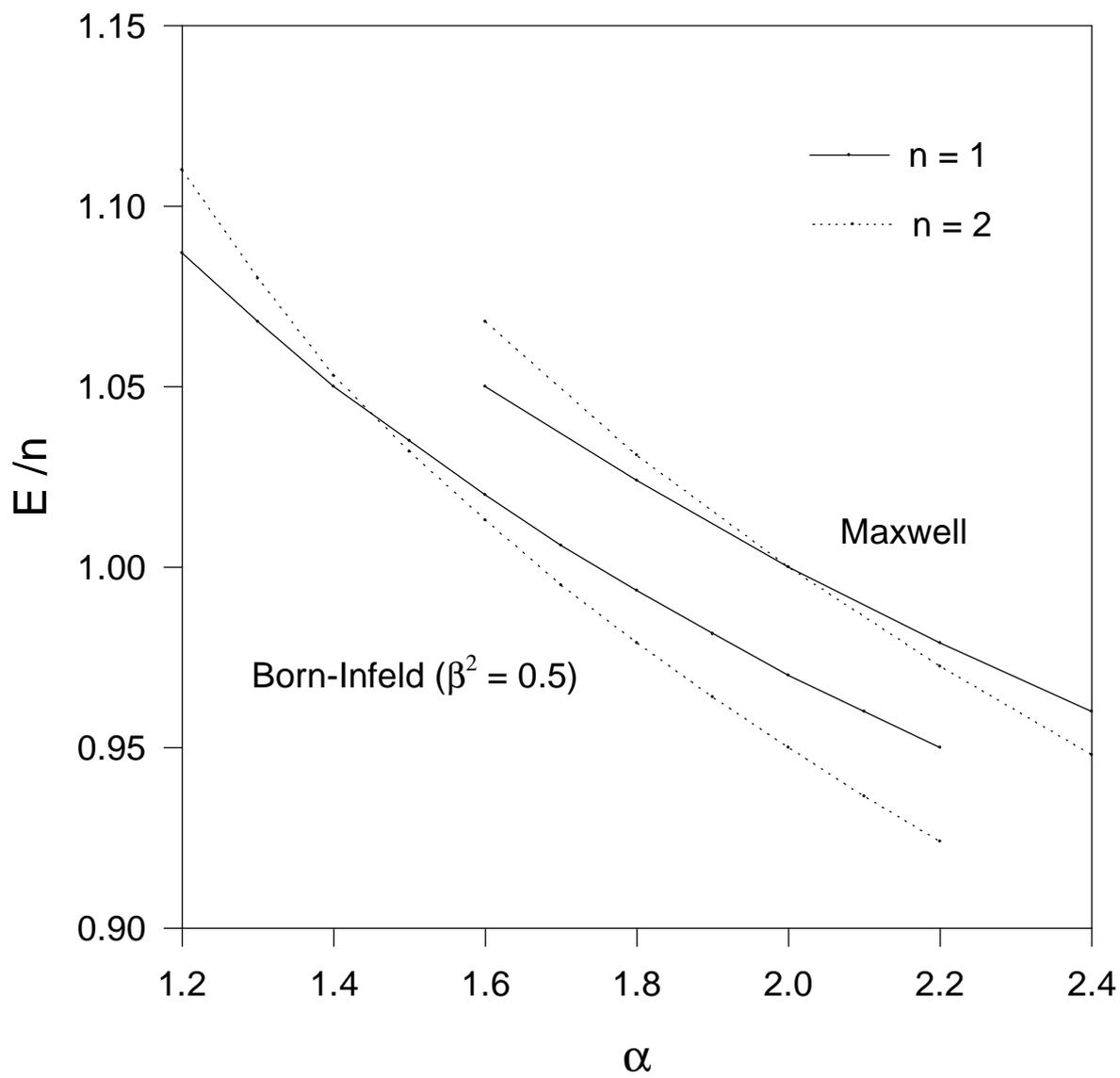
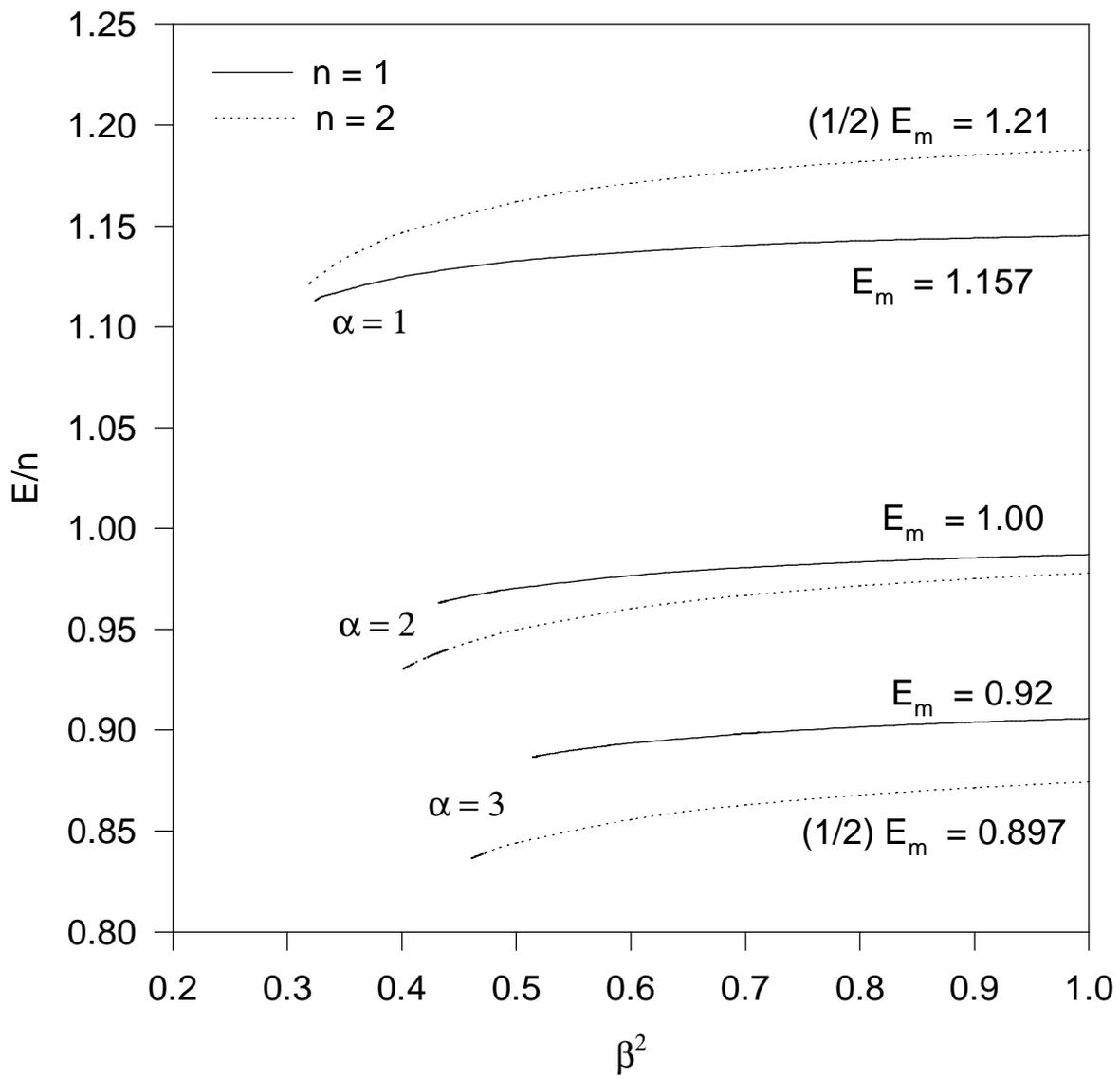
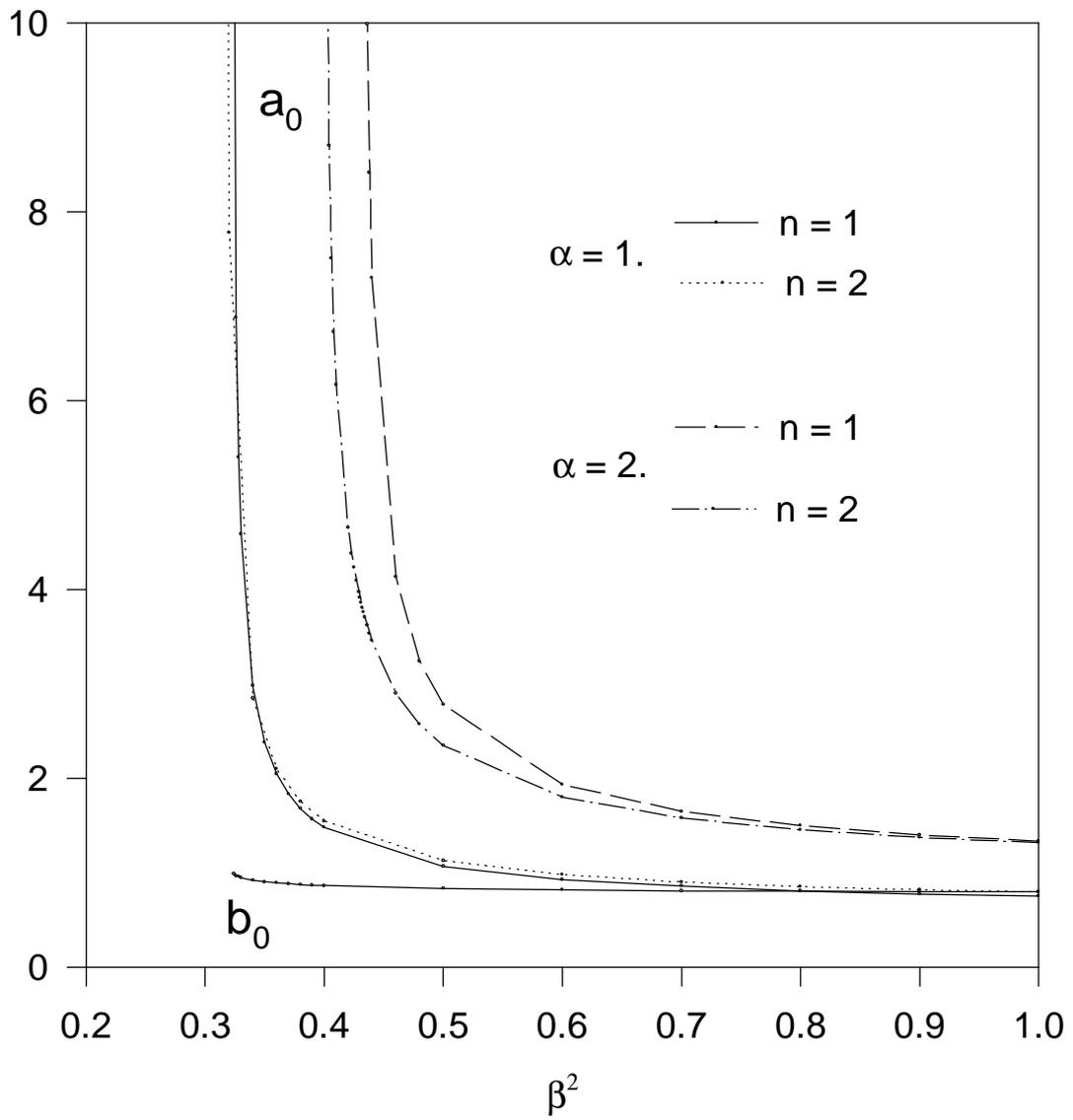
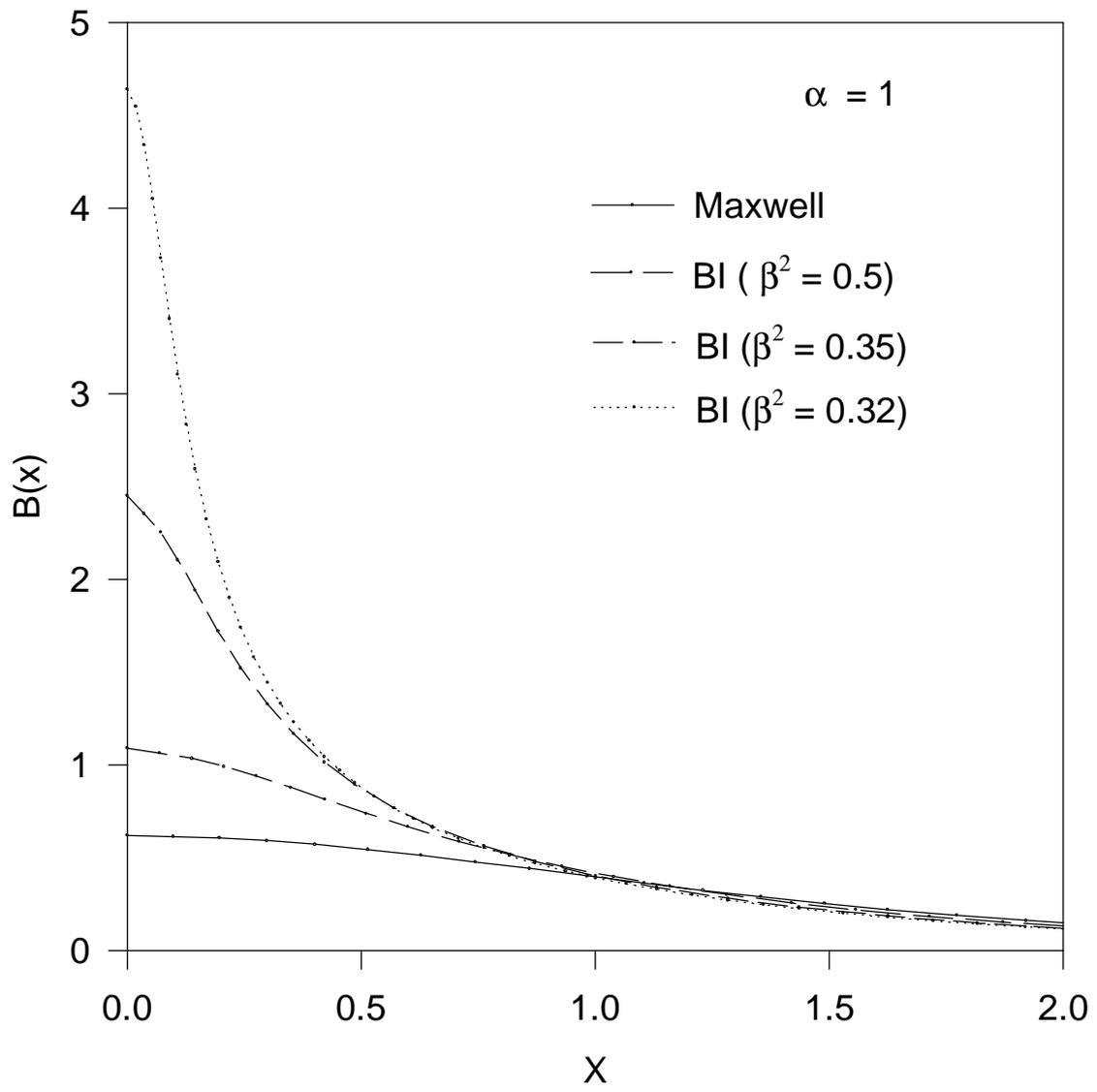
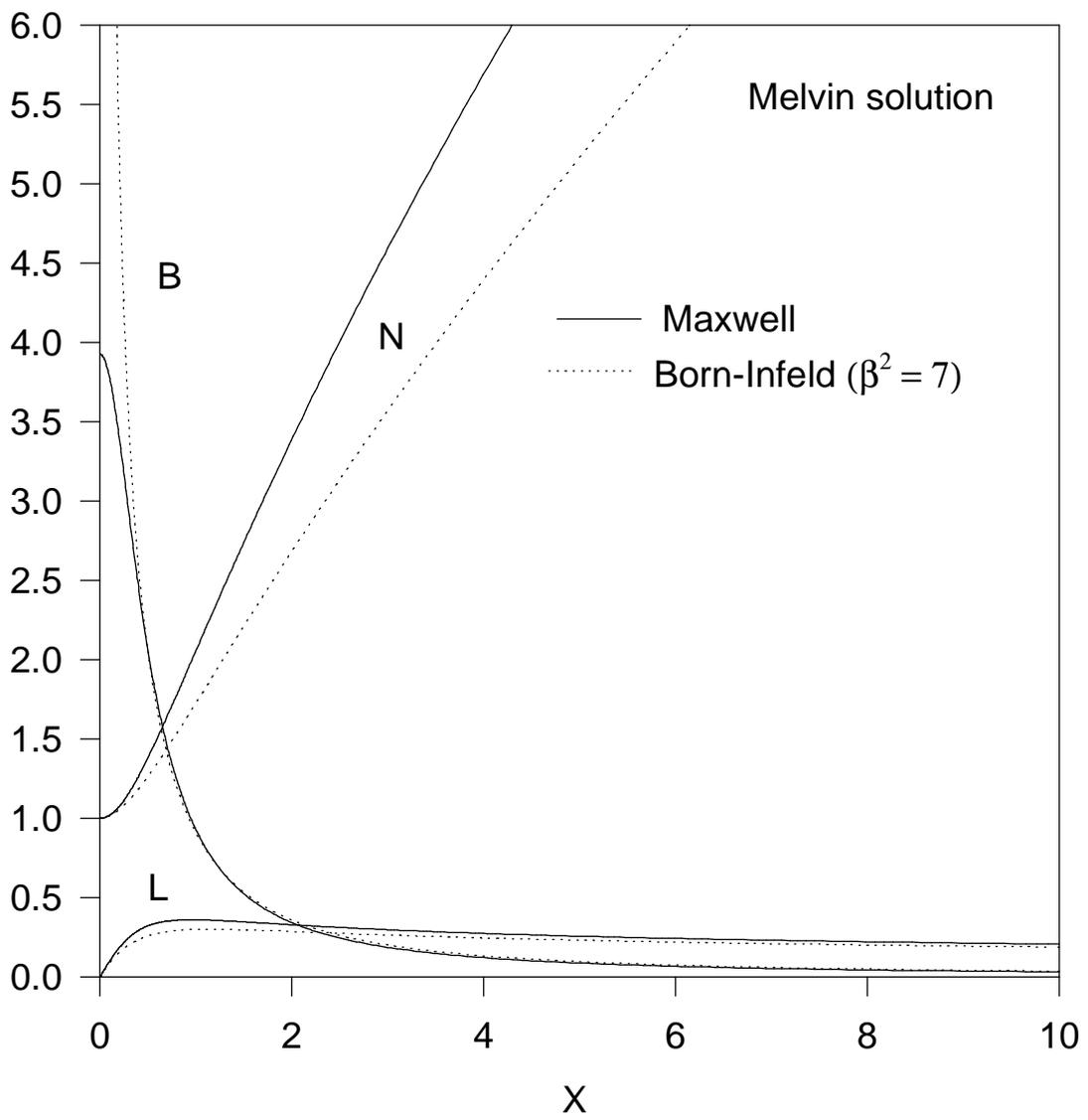
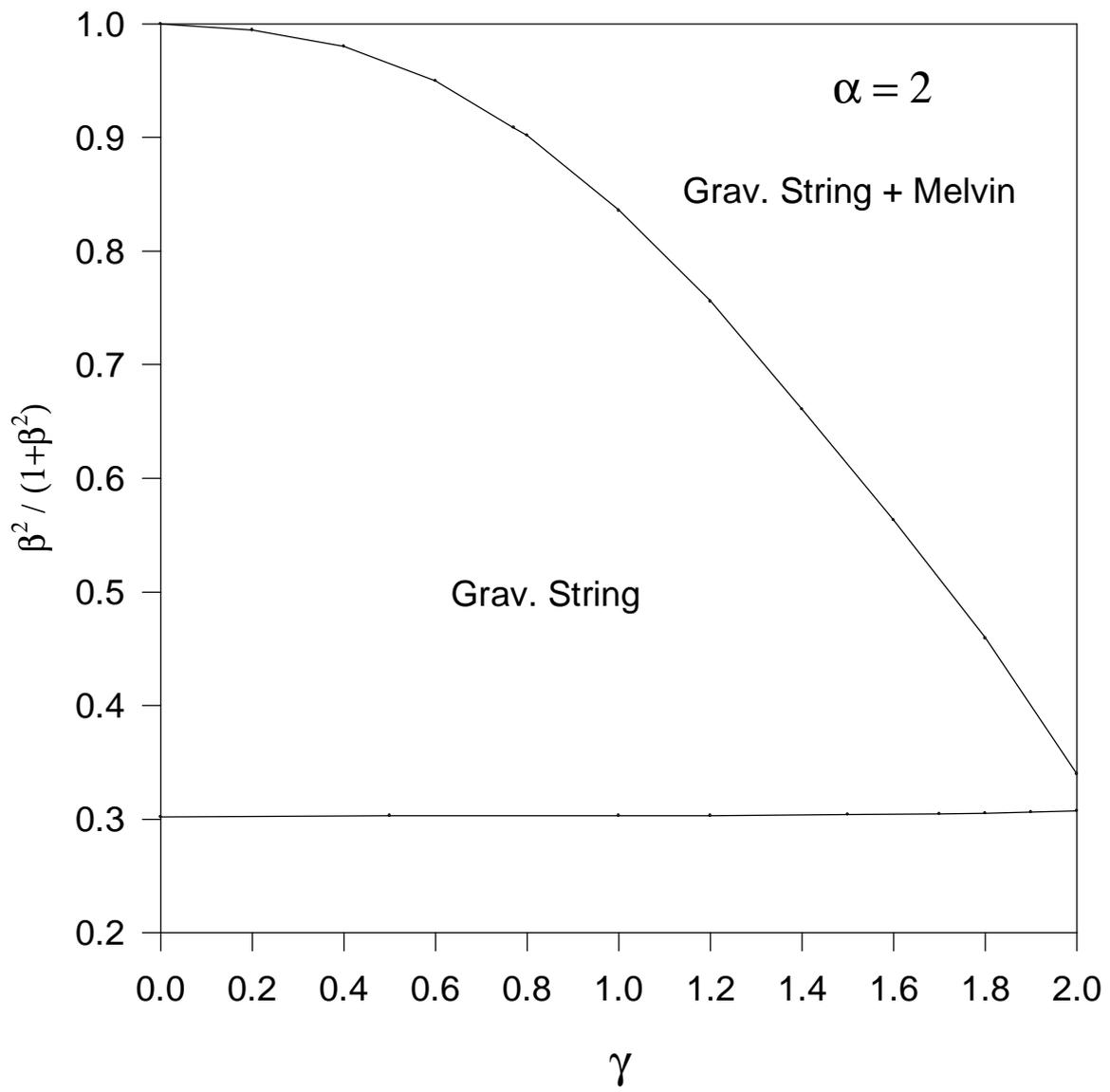
\end{document}